\documentclass[a4paper]{article}
\usepackage{epsfig,amssymb,amsmath,cite}

\def\beq{\begin{eqnarray}}
\def\eeq{\end{eqnarray}}
\def\bsp{\begin{split}}
\def\esp{\end{split}}

\newcommand{\mb}[1]{{\mathbb #1}}

\newcommand{\mbold}[1]{\mbox{\boldmath{\ensuremath{#1}}}}

\begin{document}

\title{{\bf Vacuum Plane Waves in 4+1 D \\ and \\ 
Exact solutions to Einstein's Equations in 3+1 D}}
\author{\textbf{Sigbj\o rn Hervik}\thanks{S.Hervik@damtp.cam.ac.uk}
\\ \\
 DAMTP,\\
Centre for Mathematical Sciences,\\
Cambridge University, \\
 Wilberforce Rd.,\\
Cambridge CB3 0WA, UK}
\date{\today}
\maketitle
\begin{abstract}
In this paper we derive  homogeneous vacuum plane-wave
solutions to Einstein's field equations in 4+1
dimensions. The solutions come in five different types of which three
generalise the vacuum plane-wave solutions in 3+1 dimensions to the
4+1 dimensional case. By doing
a Kaluza-Klein reduction we obtain  solutions to
the Einstein-Maxwell equations in 3+1 dimensions. The solutions generalise the
vacuum plane-wave spacetimes of Bianchi class B to
the non-vacuum case and describe spatially homogeneous
spacetimes containing an extremely tilted fluid. Also, using a similar reduction
we obtain 3+1 dimensional solutions to the Einstein equations with a scalar field. 
\end{abstract}

\section{Introduction}

Recently a full list of spatially homogeneous spacetimes in 4+1
dimensions was given \cite{sig1}. Also, for the spacetimes with simply
transitive spatial hypersurfaces, all the equations of motion were written down. In this paper we
will use these 
equations and  solve them for an important class of spacetimes:
\emph{vacuum plane-wave spacetimes possessing a covariantly constant
null Killing vector}. These spacetimes -- which have been  studied in
the literature lately (see e.g. \cite{Blau,MR}) -- have
many interesting properties. For example, the covariantly constant null vector $k^{\nu}$  satisfies
\beq
C_{\alpha\beta\mu\nu}k^{\nu}=0,\quad k^{\nu}k_{\nu}=0,
\eeq
where $C_{\alpha\beta\mu\nu}$ is the Weyl tensor; thus they are also 
algebraically  special\footnote{The 4+1 dimensional plane-wave are
also algebraically special; of type 1111 according to De Smet's scheme
\cite{DeSmet}. However, it should be noted that this classification
needs further refinement due to some pathologies.}\cite{siklos}. 

In 3+1 dimensional cosmology algebraically special 
solutions usually play a particular role; they are equilibrium points,
or sometimes even attractors, for more general classes of solutions \cite{BS1,BS,DynSys}. In particular, the vacuum plane-wave
solutions of dimension 3+1 are
attractors for non-tilted\footnote{For tilted fluids the situation is
still unsettled, but for a recent work on the futures of tilted
Bianchi models, see \cite{BHtilted}.} non-inflationary models of Bianchi class
B. Hence, general solutions of Einstein's equations of Bianchi class B
containing non-tilted non-inflationary perfect fluids 
can be approximated
with  vacuum plane-wave solutions at late times. For higher
dimensional models the situation is still unsettled as no stability
analysis has been done to date. However, if plane-wave solutions in
higher dimensions play
the same role as they do in 3+1 dimensional models then that would
definitely increase the physical relevance of spacetimes of 
plane-wave type. This paper will provide a first step in such an
analysis; we will identify them and describe them within the framework set
in \cite{sig1}. A stability analysis for 4+1 dimensional vacuum
plane-waves should then be straightforward (although it might be
somewhat lengthy). 

A different motivation for considering higher dimensional models comes
from string theory \cite{Pol,Johnson}. String theory requires higher
dimensional models 
of space-time and plane-waves spacetimes are perhaps the simplest
spacetimes not being maximally symmetric. Yet, despite the fact that they
are not maximally symmetric, the plane-wave
spacetimes  still admit supersymmetry which makes them so interesting for string theorists. Furthermore, the fact that all spacetimes have a plane
wave as a limit \cite{Penrose} has caught many theorists
attention lately due to its implications for the AdS/CFT
correspondence \cite{BMN}\footnote{It has also recently been suggested that
higher dimensional models may explain the  acceleration of
the universe observed at the present time \cite{TW}.}.

According to Kaluza-Klein theory \cite{kaluza,klein1,klein2,KaluzaKlein},  the
assumption of an extra small spatial dimension has an interesting
consequence. From a 3+1 dimensional perspective two new fields will
arise; a two-form
field ${\bf F}$ -- which can be interpreted as an electromagnetic field
tensor -- and a scalar field $\phi$. The scalar essentially measures the
size of the extra dimension -- which has to be really small in order not to be detected experimentally --  and couples to the kinetic
part of the field ${\bf F}$. This
field is often called the dilaton and in the special case where the
dilaton is constant, the theory reduces to ordinary Einstein-Maxwell
theory.\footnote{A light-like reduction, similar to the Kaluza-Klein
reduction, of these plane-wave solutions give also some interesting
results \cite{lightlike}.} 

In this paper we show that this reduction is highly effective when it
comes to generating solutions to 3+1 Einstein gravity. For
 Bianchi 
type models containing a tilted perfect fluid, there are not many
exact self-similar solutions 
known. However, by  using  insight and
knowledge of plane-wave solutions we can obtain vacuum plane-wave solutions in
4+1 dimensions. Upon reduction these solutions turn out to
correspond to  Bianchi type models of class
B containing an extremely tilted fluid. The
solutions describe  plane-wave solutions in 3+1
dimensions containing a null electromagnetic field \cite{AS}. Even
though these solutions have been found before, the reduction
procedure provides us with
a geometrical interpretation of their nature; these plane-wave spacetimes
with a null electromagnetic field are nothing but vacuum plane-waves
in one dimension higher. By studying the
Einstein-Maxwell equations, this interpretation is well hidden and
 therefore not easily recognizable. Similarly, we will also consider a
different 
special case of the Kaluza-Klein reduction. Assuming that the electromagnetic
field vanishes, the Kaluza-Klein reduction results in
Einstein gravity with a scalar field. The 3+1 dimensional scalar field
solutions  are
conformally equivalent to Bianchi class B spacetimes.

This paper is organised as follows. In section \ref{eqs} we
review some properties of homogeneous plane-wave spacetimes  and
use these to simplify the equations of
motion. In section 
\ref{planewave} we list all the solutions obtained which describes
vacuum plane-wave solutions in 4+1 dimensions. Then in section
\ref{KKreduction} we do a Kaluza-Klein reduction of the
spacetime. This leads to  exact solutions for the Einstein-Maxwell equations describing
plane-wave spacetimes of Bianchi class B with an extremely tilted
fluid and scalar field spacetimes conformally equivalent  to Bianchi type B
spacetimes.

\section{Spatially homogeneous spacetimes}
\label{eqs}
Our focus will be on spatially homogeneous spacetimes, thus we will assume
that there exist a group $G$ acting simply transitive  on the 
spatial hypersurfaces $\Sigma_t$. Following a previous paper
\cite{sig1}\footnote{For a different approach to the simply transitive
models, see \cite{Christodoulakis:2002hs,Christodoulakis:2001mg}.}, which was
based on a procedure by Ellis and MacCallum \cite{EM} in the 3+1
dimensional case, we write the spacetime (at least locally)  as a warped product 
\beq
M=\Sigma_t\times \mb{R}.
\eeq
The time-direction ${\bf u}$ can be chosen to be orthogonal to the surfaces of
transitivity. Choosing a spatial vierbein\footnote{We will use the
notation where Latin indices run over the spatial hypersurfaces while
Greek indices run over the full spacetime manifold.} ${\bf e}_a$, we can form
an orthonormal frame in spacetime $\{ {\bf e}_{\mu} \}= \{ {\bf
u},{\bf e}_a \}$. A particular useful choice of frame, is to
choose $\{ {\bf e}_a \}$ to be a \emph{left invariant frame}. If the surface
of transitivity is spanned by the Killing vectors ${\mbold\xi}_b$,
then a left invariant frame is obtained by Lie transport
\beq
\pounds_{{\mbold\xi}_b}{\bf e}_a=[{\mbold\xi}_b,{\bf e}_a]=0.
\eeq
Note that we also have 
\beq
[{\mbold\xi}_b,{\bf u}]=0.
\eeq
The corresponding dual one-forms ${\mbold\omega}^{\mu}$ obey
\beq
{\bf d}{\mbold\omega}^k=-\frac
12C^k_{~ij}{\mbold\omega}^i\wedge{\mbold\omega}^j
\eeq
where the structure constants $C^k_{~ij}$ are constants on each orbit
of transitivity.

We define the volume expansion tensor of the unit normal ${\bf u}$,
equivalently the extrinsic curvature tensor $\theta^{\mu}_{~\nu}$ of
the spatially homogeneous hypersurfaces, by
\beq
u_{\mu;\nu}=\theta_{\mu\nu}.
\eeq
One can easily check that this tensor is symmetric and purely
spatial. We split the expansion tensor into a trace part and a
trace-free part
\beq
\theta_{ab}=\frac 14\theta h_{ab}+\sigma_{ab},
\eeq
where $h_{ab}$ is the metric on the four-surfaces $\Sigma_t$,
$\theta=\theta^{\mu}_{~\mu}$ is the volume expansion scalar, and
$\sigma_{ab}$ is the shear tensor. 

Using this orthonormal frame formalism, one can derive all the
equations of motion. These equations are derived in \cite{sig1} and listed in
the generic case. 

The structure constants $C^k_{~ij}$ determine, via their Lie
algebra,  the isometry group of
the spacetime. Hence, using one of the four-dimensional real Lie
algebras, we can construct a spatially homogeneous spacetime
\cite{sig1}. 
For the plane-wave spacetimes, the most relevant Lie Algebras turn out
to be\footnote{We will use
the notation in Patera \textit{et al} \cite{PSW, PW} for the enumeration of the
4-dimensional Lie algebras. See also MacCallum's report \cite{MacC}.} $A_{4,n}$
where $n=2,3,4,5,6$. The reason for this is, as we shall see later,
that they are all non-unimodular rank 0 Lie algebras. 
Rank 0 means that these algebras have a 3
dimensional abelian Lie subalgebra. Thus they have three linearly
independent vectors ${\bf X}_A$, $A=1,2,3$ for which all the
commutators vanish identically
\beq
[{\bf X}_A,{\bf X}_B]=0.
\eeq
This enables us to write the structure constants as a matrix
\beq
C^A_{~B4}=\Theta^A_{~B}.
\eeq
For all of the rank 0 Lie algebras, except for  $A^{pq}_{4,6}$, this
matrix can, with an appropriate orientation of the orthogonal frame, be 
put onto an upper-triangular form. In Table \ref{structureconstants}
the matrix $\Theta^{A}_{~B}$ (in the particular gauge mentioned) are given for the types $A_{4,2}$ to
$A_{4,5}$. The various invariant properties of
this matrix determines the group type. The type $A^{pq}_{4,6}$ has two 
complex conjugate eigenvalues for the matrix $\Theta^A_{~B}$, but can
always be put onto the form 
\beq
A^{pq}_{4,6}:\qquad\Theta^A_{~B}=\begin{bmatrix} 
\Theta^1_{~1} & \Theta^1_{~2} & \Theta^1_{~3} \\
0 & \Theta^2_{~2} & \Theta^2_{~3} \\
0 & \Theta^3_{~2} & \Theta^2_{~2} \end{bmatrix},\qquad
\Theta^A_{~A}=3a
\eeq
where $p\Theta^1_{~1}=q\Theta^2_{~2}$ and
$\left(\Theta^2_{~2}\right)^2=-q^2\Theta^3_{~2}\Theta^2_{~3}$. 

The Lie algebra is \emph{unimodular}\footnote{A Lie group where every
left-invariant vector field preserves volume is called unimodular.} if
Tr$(\Theta^A_{~B})=0$, and 
\emph{non-unimodular} if Tr$(\Theta^A_{~B})\neq 0$. For non-unimodular
Lie algebras we can define the vector 
\beq
a_i\equiv \frac 13 C^j_{~ji},
\eeq
which, according to the contracted Jacobi identity, obeys
\beq
a_iC^i_{~jk}=0. 
\eeq
We will see that
there exists only non-trivial plane-wave solutions for non-unimodular
Lie algebras. 

\begin{table}
\centering 
\begin{tabular}{|c|c|l|}
\hline
Type &  & Restrictions on $\Theta^A_{~B}$ \\
\hline \hline
$A^{pq}_{4,5}$ & & $\Theta^1_{~1}>0,~
p\Theta^1_{~1}=\Theta^2_{~2},~q\Theta^1_{~1}=\Theta^3_{~3}$ \\
& $p=1$ & $\Theta^1_{~2}=0$ \\
& $p=q$ & $\Theta^2_{~3}=0$ \\
& $p=q=1$ & $\Theta^1_{~2}=\Theta^2_{~3}=\Theta^1_{~3}=0$ \\
\hline
$A^{p}_{4,2}$ & & $\Theta^2_{~2}=\Theta^3_{~3}\neq
0,~\Theta^1_{~1}=p\Theta^2_{~2} $ \\
 & $p=1$ & $\Theta^1_{~2}=0$\\
\hline
$A_{4,3}$ & & $\Theta^2_{~2}=\Theta^3_{~3}=0,~ \Theta^1_{~1}>0$\\
\hline 
$A_{4,4}$ & & $\Theta^1_{~1}=\Theta^2_{~2}=\Theta^3_{~3}=a$ \\
\hline
\end{tabular}
\[
\Theta^{A}_{\phantom{A}B}=\begin{bmatrix} 
\Theta^1_{~1} & \Theta^1_{~2} & \Theta^1_{~3} \\
0 & \Theta^2_{~2} & \Theta^2_{~3} \\
0 & 0 & \Theta^3_{~3} \end{bmatrix},\qquad \Theta^A_{~A}=3a.
\]
\caption{Structure constants for the Lie algebras $A_{4,2}$ to
$A_{4,5}$.}
\label{structureconstants}
\end{table}

\subsection{Tracking down the plane waves}
Homogeneous plane-wave spacetimes possessing a covariantly constant
null Killing 
vector are spaces with
higher symmetry than usual. We can use
this fact to simplify the equations of motion and find a large number of plane-wave solutions. In fact, the
technique  leads us to three-parameter families of plane-wave
solutions for each of the non-unimodular Lie algebras of rank 0
discussed in the previous section. 

The plane-wave spacetime possesses three commuting Killing vectors
${\bf X}_A$ spanning the wave front. Thus in a suitable chosen frame, the three
left-invariant vectors ${\bf e}_A$ will also be commuting. The fourth
left-invariant vector ${\bf e}_4$, has to be orthogonal to ${\bf
e}_A$ by the Jacobi identity and parallel to the  vector $a_i$. We
transform the orthonormal frame into 
a frame with two  null vectors by the
transformation 
\beq
\begin{bmatrix}
{\bf e}_+ \\ {\bf e}_- \end{bmatrix} = \frac{1}{\sqrt{2}}
\begin{bmatrix}
1 & 1 \\ -1 & 1 \end{bmatrix} \cdot \begin{bmatrix}
{\bf e}_0 \\ {\bf e}_4 \end{bmatrix}, \quad 
\begin{bmatrix}
{\mbold\eta}^+ \\ {\mbold\eta}^- \end{bmatrix} = \frac{1}{\sqrt{2}}
\begin{bmatrix}
1 & 1 \\ -1 & 1 \end{bmatrix} \cdot \begin{bmatrix}
{\mbold\omega}^0 \\ {\mbold\omega}^4 \end{bmatrix}.
\label{nulltrafo}\eeq
In this frame the metric can be written 
\beq
ds^2=2{\mbold\eta}^+{\mbold\eta}^-+\delta_{AB}{\mbold\omega}^A{\mbold\omega}^B
\eeq
For the plane-wave spacetime there does also exist a null Killing
vector which is orthogonal to the three commuting Killing vectors
spanning the wave front. Hence, ${\mbold\xi}=e^{\chi}{\bf e}_-$, where
$\chi$ is a scalar. The wave front is spanned by the three Killing
vectors ${\bf X}_A$, so ${\mbold\xi}$ must be constant along the wave
front. This implies that ${\bf X}_A(\chi)=0$. Furthermore, we require
that the null Killing
vector must commute with the left-invariant vectors ${\bf e}_A$; i.e.
\beq
[{\mbold\xi},{\bf e}_A]=0.
\eeq
Inserting the transformation eq. (\ref{nulltrafo}) implies the
following restrictions on the structure constants of the orthonormal
frame:
\beq
C^{\mu}_{~0A}=C^{\mu}_{~4A}.
\eeq
The requirement that the Killing vector ${\mbold\xi}$ is covariantly
constant, together with the orthogonality condition imply that 
\beq
C^{0}_{~0A}=C^{0}_{~4A}=C^{4}_{~0A}=C^{4}_{~4A}=C^4_{~AB}=C^0_{~AB}=0.
\eeq
Hence, we get the following relation between the volume expansion
tensor and the spatial structure constants
\beq
-\theta^{A}_{~B}+\Omega^A_{~B}=C^A_{~4B},
\label{eq:commrelation}\eeq
Here, $\Omega_{ab}$ is the angular velocity in the $ab$-plane of a
Fermi-propagated axis with respect to the triad ${\bf e}_a$.
By taking the trace of the above equation, we get
\beq
\theta-\theta^4_{~4}=3a.
\eeq
We also have
\beq
C^A_{~BC}=0,\quad a_i=a\delta^4_{~i},
\eeq
so the only non-zero commutators are therefore
\beq
C^a_{~04},\quad C^A_{~B4}\equiv\Theta^A_{~B},\quad \Theta^A_{~A}=3a.
\label{eq:remcomm}\eeq
Thus we can restrict ourselves to study the types of rank 0 which we
discussed in the previous section. Combining
eqs. (\ref{eq:commrelation}) and (\ref{eq:remcomm}),
 the expansion tensor, the
rotation tensor and the structure constants have to relate via
\beq
\theta^A_{~B}-\Omega^A_{~B}&= &\Theta^A_{~B} 
\eeq
In solving the equations of motion, we have put $\Theta^A_{~B}$ onto an upper-triangular
form as explained earlier. This can be done for all the rank 0 types,
except when 
$\Theta^A_{~B}$ has two complex conjugate eigenvalues. When  $\Theta^A_{~B}$ is
of this form, the rotation tensor $\Omega^A_{~B}$ can easily be
expressed in terms of the shear components and thus eliminated from
the equations of motion. 

\section{Vacuum plane-wave solutions}
\label{planewave}

Using the above procedure, the equations of motion reduce drasticly,
and the question of finding plane-wave solutions basically reduces to
solving a very simple set of  differential equations. First of all,
the 5D Jacobi identity and the $R_{0a}$-equations (see eqs. (22) and
(29) in \cite{sig1}) imply the constraints $\theta^4_{~A}=\Omega^4_{~A}=0$. The shear
equations, using the constraint equations,  all reduces to the same
form,
\beq
\dot{\sigma}_{AB}+\theta \sigma_{AB}-3a\sigma_{AB}=0.
\eeq
The equation for $a$ is
\beq
\dot{a}+\frac 14\theta a+\sigma_{44}a=0.
\eeq
Hence, in principle, there are only two types of equations that needs
to be solved, subject to the constraint equations. 

In the following
the solutions 
for the rank 0 non-unimodular Lie algebras are given. The metrics
correspond to vacuum plane-wave solutions where the gravitational
wave propagates in the $w$-direction. They all possess the
covariantly constant null Killing vector 
\beq
{\mbold\xi}=e^{-t+w}\left(\frac{\partial}{\partial
t}-\frac{\partial}{\partial w}\right).
\eeq

These spacetimes are also self-similar; hence,
they possess a homothetic vector field. This homothetic vector field is
given by
\beq
{\mbold\xi}_H=\frac{\partial}{\partial
t}-\frac{\partial}{\partial w}+x\frac{\partial}{\partial
x}+y\frac{\partial}{\partial y} +z\frac{\partial}{\partial z}
\eeq
as can be easily verified.

\subsection{Type $A^{pq}_{4,5}$}
Let $\beta_+, \ \beta_-, \ Q_1,\ Q_2,\ Q_3$ be free parameters such
that 
\beq 8(\beta_+^2+ \beta_-^2)+\frac{2}{3}(Q_1^2+Q_2^2+Q_3^2) \leq 1.
\eeq
A five parameter set of plane-wave solutions can be given by (in upper
triangular form)
\beq
ds^2 =&& e^{2t}(-dt^2+dw^2) + e^{2s(w+t)}\nonumber \\
&\times&\bigg[e^{-4\beta_+(w+t)}\left(dx+\frac{Q_1}{P_1}e^{P_1(w+t)}dy+\frac{Q_1Q_3+P_3Q_2}{P_3P_2}e^{P_2(w+t)}dz\right)^2
\nonumber \\
&+&
e^{2(\beta_++\sqrt{3}\beta_-)(w+t)}\left(dy+\frac{Q_3}{P_3}e^{P_3(w+t)}dz\right)^2\nonumber
\\
&+&e^{2(\beta_+-\sqrt{3}\beta_-)(w+t)}dz^2\bigg]
\eeq
where
\beq
s(1-s)&=& 2(\beta_+^2+
\beta_-^2)+\frac{1}{6}(Q_1^2+Q_2^2+Q_3^2)\nonumber \\
P_1 &=& 3\beta_++\sqrt{3}\beta_- \nonumber \\ 
P_2 &=& 3\beta_+-\sqrt{3}\beta_- \nonumber \\ 
P_3 &=& -2\sqrt{3}\beta_-.
\label{eq:Ps}\eeq
The group parameters are related to these parameters as follows
\beq
p &=& \frac{s + (\beta_++\sqrt{3}\beta_-)}{s+(\beta_+-\sqrt{3}\beta_-)} \nonumber \\
q &=& \frac{s-2\beta_+}{s + (\beta_+-\sqrt{3}\beta_-)}.
\eeq
The requirement that $p\neq 1$ leads to $P_3\neq 0$ (see Table \ref{structureconstants}). Similarly, $q\neq
1$ and $p\neq q$ lead to $P_2\neq 0$ and $P_1\neq 0$ respectively. If
$P_i=0$ then we have to set $Q_i=0$ first. 

Hence, two of the parameters define what group type the
spacetime belongs to. For each group type we have a three-parameter family
of plane-wave solutions (roughly given by the parameters $Q_1,\ Q_2$ and
$Q_3$). 

In the unimodular limit (where the trace of the commutators is zero)
we have $s\longrightarrow 0$. The only way to obtain $s=0$ is when
$\beta_{\pm}=Q_i=0$. Thus these plane-wave solutions approaches the
Minkowski spacetime in this limit. 

We can recover the VI$_h\oplus{\mb{R}}$ plane waves by requiring
$q=0$. Thus we set $s=2\beta_+$ and the metric simplifies  to 
\beq
ds^2 =&& e^{2t}(-dt^2+dw^2)  \nonumber \\
 &+& \left(dx+\frac{Q_1}{P_1}e^{P_1(w+t)}dy+\frac{Q_1Q_3+P_3Q_2}{P_3P_2}e^{P_2(w+t)}dz\right)^2
\nonumber \\
&+&e^{(3s+\sqrt{3}\beta_-)(w+t)}\left(dy+\frac{Q_3}{P_3}e^{P_3(w+t)}dz\right)^2
\nonumber \\&+ &e^{(3s-\sqrt{3}\beta_-)(w+t)}dz^2.
\eeq
Hence, in general we can have VI$_h\oplus{\mb{R}}$ plane-wave solutions
where the extra dimension is tilted. This is exactly what will lead to
the electromagnetic field when we reduce this metric to the 3+1 case. 
 To recover the  VI$_h\oplus{\mb{R}}$ plane
waves where the extra dimension is trivially added, we must put $Q_1=Q_2=0$. 

\subsection{Type $A^p_{4,2}$}
Let $\beta_+, \ Q_1,\ Q_2,\ Q_3$ be free parameters such
that 
\beq 8\beta_+^2+\frac{2}{3}(Q_1^2+Q_2^2+Q_3^2) \leq 1.
\eeq
A four parameter set of plane-wave solutions can be given by (in upper
triangular form)
\beq
ds^2 =&& e^{2t}(-dt^2+dw^2) + e^{2s(w+t)}\nonumber \\
&\times&\bigg[e^{-4\beta_+(w+t)}\left(dx+\frac{Q_1}{P_1}e^{3\beta_+(w+t)}dy+\left[A+B(w+t)\right]e^{3\beta_+(w+t)}dz\right)^2
\nonumber \\
&+&
e^{2\beta_+(w+t)}\left(dy+{Q_3}(w+t)dz\right)^2\nonumber
\\
&+&e^{2\beta_+(w+t)}dz^2\bigg]
\eeq
where $s$ and $P_i$ are given in eq. (\ref{eq:Ps}) with $\beta_-=0$, and
\beq
A &=& \frac{3\beta_+Q_2-Q_1Q_3}{9\beta_+^2} \nonumber \\
B &=& \frac{Q_1Q_3}{3\beta_+}.
\label{eqAB}\eeq
The group parameter is given by
\beq
p=\frac{s-2\beta_+}{s+\beta_+}.
\eeq
In the limit where $p\rightarrow 0$ we end up in the decomposable
case IV$\oplus \mb{R}$. Also in this case the plane-wave solutions
describe in general a 
tilted extra dimension. 

\subsection{Type $A_{4,3}$}
Plane-wave solutions for the Lie algebra type $A_{4,3}$ can be
obtained by taking the $p\longrightarrow\infty$ limit of
$A_{4,2}^p$. In this limit we get $\beta_+=-s$ and thus the metric can
be written as
\beq
ds^2 =&& e^{2t}(-dt^2+dw^2) \nonumber \\
&+&e^{6s(w+t)}\left(dx+\frac{Q_1}{P_1}e^{-3s(w+t)}dy+\left[A+B(w+t)\right]e^{-3s(w+t)}dz\right)^2
\nonumber \\
&+&\left(dy+{Q_3}(w+t)dz\right)^2
+dz^2
\eeq
where 
\beq
s=\frac 16\left(1\pm\sqrt{1-2(Q_1^2+Q_2^2+Q_3^2)}\right),
\eeq
and $A,B$ are given in eq. (\ref{eqAB}) with $\beta_+=-s$. 

\subsection{Type $A_{4,4}$}
The parameters $Q_i$ are free parameters and $s$ is given in eq. (\ref{eq:Ps})
with $\beta_{\pm}=0$. There is a three-parameter set of plane-wave
solutions given by 
\beq
ds^2 =&& e^{2t}(-dt^2+dw^2) + e^{2s(w+t)}\nonumber \\
&\times&\bigg[\left(dx+{Q_1}(w+t)dy+(w+t)\left[Q_2+\frac{Q_1Q_3}{2}(w+t)\right]dz\right)^2
\nonumber \\
&+&\left(dy+{Q_3}(w+t)dz\right)^2 +dz^2\bigg].
\eeq
In the limit where the trace of the structure constants go to zero,
$a\longrightarrow 0$, and thus $s\longrightarrow 0$. Hence, we have
$Q_i=0$ and we recover flat five-dimensional Minkowski space. $Q_i=0$
yields also the possibility $s=1$. In that case the above metric
reduces to the five-dimensional Milne universe.

\subsection{Type $A_{4,6}^{pq}$}
Let $\beta_{+}, \omega, \beta, Q_1, Q_2$ be free parameters and let $s$ be
given by
\beq
s(1-s)=2\beta_+^2+\frac 23\omega^2\sinh^22\beta+\frac 16(Q_1^2+Q_2^2).
\eeq
Define also the two one-forms
\beq
{\mbold\omega}^2&=& \cos[\omega(w+t)] {\bf dy}-\sin[\omega(w+t)]
{\bf dz} \nonumber \\
{\mbold\omega}^3&=& \sin[\omega(w+t)] {\bf dy}+\cos[\omega(w+t)] {\bf dz}.
\eeq
The plane-wave solutions of type $A_{4,6}^{pq}$ can now be written
\beq
ds^2 =&& e^{2t}(-dt^2+dw^2)+e^{2s(w+t)} \nonumber \\
&\times&
\bigg[e^{-4\beta_+(w+t)}\left\{dx+e^{3\beta_+(w+t)}\left(q_1e^{-\beta}{\mbold\omega}^3-q_2e^{\beta}{\mbold\omega}^2\right)\right\}^2
\nonumber \\
&+&e^{2\beta_+(w+t)}\left\{
e^{-2\beta}\left({\mbold\omega}^2\right)^2+e^{2\beta}\left({\mbold\omega}^3\right)^2\right\}\bigg]
\eeq
where 
\beq
q_1 &=& \frac{Q_1\omega+3\beta_+Q_2e^{2\beta}}{\omega^2+9\beta_+^2}
\nonumber \\
q_2 &=& \frac{Q_2\omega-3\beta_+Q_1e^{-2\beta}}{\omega^2+9\beta_+^2}.
\eeq
The group parameters are related to these constants via
\beq
p &=& \frac{\beta_+(s-2\beta_+)}{\omega(s+\beta_+)}\nonumber \\
q &=& \frac{\beta_+}{\omega}.
\eeq
In the VII$_h\oplus\mb{R}$ limit, we have $s=2\beta_+$. The spacetime
has in general a tilted extra dimension and we will recover the
VII$_h$ plane waves only if we simultaneously set $Q_1=Q_2=0$. 

\section{Kaluza-Klein reduction}
\label{KKreduction}
Kaluza-Klein theory  considers a five-dimensional spacetime
which possesses (at least) one spatial Killing vector ${\mbold\xi}$. If, say,
this Killing vector has the corresponding left-invariant form ${\bf
dx}$, when  the metric can be written as
\beq
ds^2=g_{\mu\nu}{\mbold\omega}^{\mu}{\mbold\omega}^{\nu}+\phi^2({\bf
dx}+A_{\mu}{\mbold\omega}^{\mu})^2.
\eeq
The Kaluza-Klein reduction can now be performed if we identify the
space under a translation in the ${\mbold\xi}$-direction. The fifth
dimension is assumed to be small and its size will in general define
the electromagnetic coupling constant. Also, the field 
\beq
{\bf A}\equiv A_{\mu}{\mbold\omega}^{\mu}
\eeq
can be interpreted as a vector potential for an electromagnetic
field. Geometrically this vector potential arises whenever the Killing
vector ${\mbold\xi}$ is not orthogonal to the reduced four-dimensional
spacetime. This vector potential is related to the electromagnetic
field strength, ${\bf F}$, in the usual manner,
\beq
{\bf F}={\bf dA}.
\eeq 
In general we will have a non-constant dilaton field $\phi$ which
couples to the electromagnetic field in a way not standard in Einstein
gravity. However, we will consider two special cases, both of which can be
interpreted within the standard Einstein gravity. 

\subsection{Exact solutions to the Einstein-Maxwell equations}
We have already a set of vacuum solutions in 4+1 dimensions. These
must correspond to a set of solutions in 3+1 dimensions to the
Einstein-Maxwell-dilatonic equations of motion; according to the
Kaluza-Klein theory, it is only a
matter of reinterpreting the solutions. In general we will obtain a
dilatonic field $\phi$ but we will first require 
$\phi=constant$. The analysis can trivially be expanded to include a
non-constant $\phi$. After reduction the solutions must
therefore be solutions to Einstein's field equations 
in 3+1 dimensions with an electromagnetic field. Hence, they are
solutions to the \emph{Einstein-Maxwell} equations! 

By inspection of the vacuum plane-wave solutions in 4+1
dimensions, we readily see that  the constraint $\phi=constant$  can be fulfilled by
requiring
\beq
s=2\beta_+
\eeq
when we are compactifying along the $\frac{\partial}{\partial x}$
Killing vector. The compactification can in principle be along any of the Killing vectors
spanning the wave-front\footnote{We also have to require that the
action of the Killing vector acts freely on spacetime, otherwise the
resulting spacetime will have orbifold-singularities.}, but we see
that up to permutations all the non-trivial solutions with
$\phi=constant$ can be obtained this way. Hence, there is no loss of
generality to assume that we compactify along $\frac{\partial}{\partial x}$. 

Apart from the requirement $s=2\beta_+$, there are no further
restrictions on the parameters. Performing the reduction on
$A^p_{4,2}$, $A^{pq}_{4,5}$ and $A^{pq}_{4,6}$ (the others lead to
trivial spacetimes) yield a Bianchi type IV, VI$_h$ and VII$_h$
spacetime, respectively. The spaces possess an electromagnetic two-form
field (renaming the $w$ direction as $x^1$) given by
\beq
{\bf F}={\bf
dA}=e^{-t}({\mbold\omega}^0+{\mbold\omega}^1)\wedge(Q_1{\mbold\omega}^2+Q_2{\mbold\omega}^3). 
\eeq
Here, we have introduced an orthonormal frame ${\mbold\omega}^{\mu}$,
so that 
\beq
ds^2=\eta_{\mu\nu}{\mbold\omega}^{\mu}{\mbold\omega}^{\nu},
\eeq
and $\eta_{\mu\nu}$ is the Minkowski metric. The solutions are 
\beq
IV: &&
\begin{cases}
{\mbold\omega}^0 = e^t{\bf dt} \\
{\mbold\omega}^1 = e^t{\bf dw} \\
{\mbold\omega}^2 = e^{s(w+t)}[{\bf dy}+Q_3(w+t){\bf dz}]\\
{\mbold\omega}^3 = e^{s(w+t)}{\bf dz}  \\
s(1-s) = \frac 14(Q_1^2+Q^2_2+Q^2_3)
\end{cases}\\
VI_h: && 
\begin{cases}
{\mbold\omega}^0 = e^t{\bf dt} \\
{\mbold\omega}^1 = e^t{\bf dw} \\
{\mbold\omega}^2 = e^{(s+b)(w+t)}[{\bf dy}-\frac{Q_3}{2b}e^{-2b(w+t)}{\bf dz}]\\
{\mbold\omega}^3 = e^{(s-b)(w+t)}{\bf dz}  \\
s(1-s) = b^2+\frac 14(Q_1^2+Q^2_2+Q^2_3)
\end{cases}\\
VII_h: &&
\begin{cases}
{\mbold\omega}^0 = e^t{\bf dt} \\
{\mbold\omega}^1 = e^t{\bf dw} \\
{\mbold\omega}^2 = e^{s(w+t)}e^{-\beta}\left\{\cos[\omega(w+t)] {\bf
dy}-\sin[\omega(w+t)] {\bf dz}\right\}\\
{\mbold\omega}^3 = e^{s(w+t)}e^{\beta}\left\{\sin[\omega(w+t)] {\bf
dy}+\cos[\omega(w+t)] {\bf dz}\right\}
  \\
s(1-s) = \omega^2\sinh^22\beta+\frac 14(Q_1^2+Q^2_2).
\end{cases}
\eeq
Note that all these solutions are self-similar, since they all possess
the homothety
\beq
{\mbold\xi}_H=\frac{\partial}{\partial
t}-\frac{\partial}{\partial w}+y\frac{\partial}{\partial y} +z\frac{\partial}{\partial z}.
\eeq
One can  readily verify that the source-free Maxwell equations
\beq
{\bf dF}=0, \qquad {\bf d}^{\dagger}{\bf F}=0,
\eeq
and the Einstein field equations are satisfied (for the specific
choice of constants $16\pi G=e=c=1$).

The type IV,
  VI$_h$,  and  VII$_h$ metrics are the same as those in
  \cite{AS}\footnote{Thanks to Professor MacCallum for bringing this
  paper to my attention.} and
generalise Harvey and Tsoubelis' \cite{HT}, Collins'\cite{Collins},
and  Lukash's  \cite{Lukash}
vacuum plane-wave solutions, respectively.\footnote{Note
that all  these solutions are given in \cite{Kramer}, but they are
all given
in the more general Brinkman form,
\[
ds^2=2dudv+H(u,x^i)du^2+dx^idx_i.
\] 
In the paper \cite{AS}, Araujo and
  Skea used these exact solutions in Brinkman form and identified the
  ones admitting a simply transitive group of isometries.} The source is an
electromagnetic field which is of a very particular type. The electric
and magnetic 
fields are (in the orthonormal frame)
\beq
E_i=e^{-t}(0,-Q_1,-Q_2),\qquad B_i=e^{-t}(0,-Q_2,Q_1).
\eeq
Thus this is a \emph{null} field where all invariants composed of the
two-form field vanish\footnote{$F_{\mu\nu}F^{\mu\nu}=0$ is, in fact, a
  direct consequence of the requirement $\phi= constant$.}
\beq
F_{\mu\nu}F^{\mu\nu}=F_{\mu\nu}\left(\star F\right)^{\mu\nu}=0.
\eeq
The energy density of the field is
\beq
\rho_{EM}=\frac 12(E^2+B^2)=e^{-2t}(Q_1^2+Q_2^2).
\eeq
These solutions describes plane-wave spacetimes with a plane-wave
electromagnetic field propagating in it. Generally they are non-vacuum
plane-wave solutions of Bianchi types IV, VI$_h$ and
VII$_h$. They interpolate between a
one-parameter\footnote{These solutions are axisymmetric and hence one
of the parameters can be gauged away by a rotation.} family of
electro-magneto-vacuum spacetimes of type V and the well known vacuum plane-wave
solutions of type IV, VI$_h$ and
VII$_h$. 

Note also that the energy-momentum tensor is given by
\beq
T_{\mu\nu}=\rho_{EM}\cdot\begin{bmatrix}
1& 1& 0 & 0\\
1& 1& 0 & 0\\
0& 0& 0 & 0\\
0& 0& 0 & 0\\
\end{bmatrix},
\label{eq:EMtensor}\eeq
and hence, describes a universe containing a fluid with extreme
tilt. According to an observer 
comoving with the homogeneous hypersurfaces, there will be a stream of
photons in the ${\bf e}_1$-direction. The  momentum-flow is therefore
null, and thus describes a fluid  with extreme tilt. 

We know that the vacuum
plane-wave solutions in 3+1 dimensions are attractor solutions for a
large class of spacetimes containing \emph{non-tilted} perfect fluids
\cite{BS}. However, what happens for spacetimes containing a tilted
fluid is still  uncertain. A recent work indicates
that tilted fluids stiffer than radiation may be important for the
late time behaviour of Bianchi cosmologies class B
\cite{BHtilted}. What is certainly 
the case is that these 
solutions provide us with  examples of type IV and VII$_h$ spacetimes which do
not asymptote a \emph{vacuum} plane-wave at late times. Hence, the
vacuum solutions of Lukash cannot be global attractors not even within the
Einstein-Maxwell field equations. 

\subsection{Exact solutions to the Einstein equations with a scalar
field}
There is another case where it is possible to interpret the solutions
after the reduction procedure classically. Consider the case  $\phi$
non-constant. Let $ds^2_5$ be a 5 dimensional vacuum 
solution and $ds^2_4$ be the reduced 4 dimensional metric; i.e. 
\beq
ds_5^2=ds^2_4+\phi^2({\bf
dx}+A_{\mu}{\mbold\omega}^{\mu})^2.
\eeq
Then, if ${\bf A}=0$, the 4 dimensional conformally related metric 
\beq
d\tilde{s}_4^2=\phi ds^2_4,
\eeq
 will satisfy the 4D Einstein equations with a free scalar
field. Alternatively, the scalar field can be thought of as a stiff
fluid with equation of state $\rho=p$. 

Again one get three cases where the metrics are all conformally
related to Bianchi type IV, VI$_h$, and VII$_h$ metrics. However, in
general the
conformal factor breaks the spatial homogeneity so the
solutions are no longer spatially homogeneous. In all the cases, the
requirement ${\bf A}=0$ implies $Q_1=Q_2=0$. Using an orthonormal frame
${\mbold\omega}^{\mu}$, the solutions are
\beq
\mathcal{C}IV: &&
\begin{cases}
{\mbold\omega}^0 = e^{\Omega(w+t)+t}{\bf dt} \\
{\mbold\omega}^1 = e^{\Omega(w+t)+t}{\bf dw} \\
{\mbold\omega}^2 = e^{s(w+t)}[{\bf dy}+Q_3(w+t){\bf dz}]\\
{\mbold\omega}^3 = e^{s(w+t)}{\bf dz}  \\
s(1+2\Omega-s) = 3\Omega^2+\frac 14Q^2_3
\end{cases}\\
\mathcal{C}VI_h: && 
\begin{cases}
{\mbold\omega}^0 = e^{\Omega(w+t)+t}{\bf dt} \\
{\mbold\omega}^1 = e^{\Omega(w+t)+t}{\bf dw} \\
{\mbold\omega}^2 = e^{(s+b)(w+t)}[{\bf dy}-\frac{Q_3}{2b}e^{-2b(w+t)}{\bf dz}]\\
{\mbold\omega}^3 = e^{(s-b)(w+t)}{\bf dz}  \\
s(1+2\Omega-s) = 3\Omega^2+b^2+\frac 14Q^2_3
\end{cases}\\
\mathcal{C}VII_h: &&
\begin{cases}
{\mbold\omega}^0 = e^{\Omega(w+t)+t}{\bf dt} \\
{\mbold\omega}^1 = e^{\Omega(w+t)+t}{\bf dw} \\
{\mbold\omega}^2 = e^{s(w+t)}e^{-\beta}\left\{\cos[\omega(w+t)] {\bf
dy}-\sin[\omega(w+t)] {\bf dz}\right\}\\
{\mbold\omega}^3 = e^{s(w+t)}e^{\beta}\left\{\sin[\omega(w+t)] {\bf
dy}+\cos[\omega(w+t)] {\bf dz}\right\}
  \\
s(1+2\Omega-s) = 3\Omega^2+\omega^2\sinh^22\beta.
\end{cases}
\label{eq:scalarsolutions}\eeq
The scalar field is given by $\varphi=-2\sqrt{3}\Omega (w+t)$, and the
energy-momentum tensor has the same form as
eq.(\ref{eq:EMtensor}). Hence, the scalar field is extremely tilted. 

These solutions can be interpreted as stiff fluid solutions which can
be generated  using the
generation techniques described in Theorem 10.2 in \cite{Kramer}. They
reduce to the ordinary vacuum plane wave when $\Omega=0$, and this is
the only case they are spatially homogeneous. There are, however,
always an abelian $\mb{R}^3$ acting simply transitive on the null
hypersurfaces $w+t=\text{constant}$. The 
metrics are all conformally related to Bianchi types of class B as
indicated in eq. (\ref{eq:scalarsolutions}). 

What role these solutions may have for the evolution of generic
cosmological models is unknown to the author. 

\section{Conclusion}
We have found 5 classes of homogeneous vacuum plane-wave solutions in 4+1
dimensions. For each non-unimodular Lie algebra of rank 0, there is a
three-parameter family of solutions to the Einstein field
equations. An open question which we have not addressed here, but
nonetheless would be very interesting to answer, is: Are these
solutions the late time attractors for a more general set of models? We
know that the 3+1 dimensional vacuum plane waves are late-time
attractors for non-tilted perfect fluid cosmologies of Bianchi class B
\cite{DynSys}. Therefore we might wonder if the 4+1 dimensional plane
waves plays the same role in 4+1 dimensional cosmology? To this date,
no such stability analysis has been done mostly because the lack of
understanding of spatially homogeneous cosmological models of
dimension 4+1. However, by the previous work \cite{sig1}, and this
work, such a stability analysis seems to be within reach. Such an analysis
will hopefully give us some hints whether 3+1 dimensional is special
or if it has the same features as higher dimensional cosmology.

By doing a Kaluza-Klein reduction, we showed explicitly that some of
these solutions 
correspond to the exact solutions of the Einstein-Maxwell equations
found by Araujo and Skea \cite{AS} using a different type of analysis. These
solutions describe spacetimes containing an extremely  tilted
fluid. The benefit 
of the Kaluza-Klein theory is therefore twofold; not only can we
obtain exact solutions in 3+1 dimensions, but in doing so we also
give a geometrical 
explanation of their nature. The obtained solutions of the
Einstein-Maxwell equations are nothing but vacuum plane-waves in
one dimension higher.  Amazingly, this sets the 4+1 dimensional
cosmology in an incredible context. The study of 4+1 dimensional
cosmology may help us to understand 3+1 dimensional cosmology.
In this context the plane-wave spacetimes in
homogeneous form have a great advantage with respect to the more commonly used
Brinkman form. 

Also 3+1 dimensional solutions with a massless scalar field were
derived using the Kaluza-Klein reduction. These solutions were not spatially
homogeneous in general but they possessed three commuting Killing vectors
acting simply transitively on  null hypersurfaces. These are
solutions that can be obtained by well known generation techniques
\cite{Kramer}.

Notwithstanding, this may indicate that cosmology in 3+1 dimensions
may simplify and be a lot more comprehensible by increasing the
dimensionality by one. Hence, the study of 4+1 dimensional models may
be more down-to-Earth than we originally might have thought. 

\section*{Acknowledgments} The author deeply acknowledges the
insightful and useful comments made by S.T.C. Siklos, J.D. Barrow and
A.A. Coley. Thanks also to James Lucietti for reading through the
manuscript and making useful comments. 
This work was funded by the Research Council of Norway and an Isaac
Newton Studentship.

\end{document}